\documentstyle[twocolumn,aps,psfig]{revtex}
\begin{document}
%\draft

\twocolumn[\hsize\textwidth\columnwidth\hsize\csname @twocolumnfalse\endcsname

\title{ Enhancement of Antiferromagnetic Correlations Induced by
Nonmagnetic Impurities: Origin and Predictions for NMR Experiments}

\author{Markus Laukamp, George Balster Martins,  
Claudio Gazza, Andr\'e L. Malvezzi,
and Elbio Dagotto}
\address{National High Magnetic Field Lab and Department of Physics, \\ Florida State University, Tallahassee,
Florida 32306,USA}
\author{Patricia M. Hansen, Alfredo C. L\'opez, and Jos\'e Riera}
\address{Instituto de F\'{\i}sica Rosario y Departamento de F\'{\i}sica, Avenida Pellegrini 250, 
2000 Rosario,
Argentina}
\maketitle

\begin{abstract}

Spin models that have been proposed to describe
dimerized chains, ladders, two dimensional antiferromagnets, and
other compounds are here studied when some spins are replaced
by spinless vacancies, such as it occurs 
by $Zn$ doping. A small percentage of vacancies rapidly destroys 
the spin gap, and their presence induces enhanced
antiferromagnetic correlations near those vacancies.
The study is performed with computational techniques which includes
Lanczos, world-line 
Monte Carlo, and the Density Matrix Renormalization Group
methods.
Since the phenomenon of enhanced antiferromagnetism is found to
occur in several models and
cluster geometries, a common simple explanation for its
presence may exist. It is argued that the resonating-valence-bond
character of the spin correlations at short distances of a large
variety of models is responsible for the presence of robust
staggered spin correlations near vacancies and lattice edges. The
phenomenon takes place regardless of the long distance properties
of the ground state, and it is caused by a ``pruning'' of the available
 spin singlets in the vicinity of the vacancies.
The effect produces a broadening of the low temperature NMR
signal for the compounds analyzed here.
This broadening
should be experimentally
observable in the structurally dimerized chain systems
 $Cu(NO_3)_2\cdot2.5H_2O$, $CuWO_4$, $(VO)_2P_2O_7$, and 
$Sr_{14}Cu_{24}O_{41}$, in ladder materials such as $Sr Cu_2 O_3$,
in the spin-Peierls systems $CuGeO_3$ and $NaV_2 O_5$,
and in several others since it is a universal effect common to a wide
variety of models and compounds.

\end{abstract}


\vskip2pc]
\narrowtext

\section{Introduction}

The theoretical and experimental study of quasi-one dimensional compounds 
continues attracting considerable attention. Among these materials a
large number of  systems which can be described by Hamiltonians where
the relevant degrees of freedom are localized spins
 have been studied using a
wide variety of techniques. Interesting spin gapped and gapless systems
and models have been identified.
In particular, compounds with
 ``ladder'' structures are currently under much 
investigation~\cite{ladder}. The excitement surrounding these 
compounds was originally triggered  by theoretical predictions
suggesting
(i) the existence of a spin gap in the undoped case, and 
(ii) a transition to a superconducting state upon hole doping.
A vast amount of experimental literature has indeed confirmed that ladder
materials, such as $Sr Cu_2 O_3$, are gapped~\cite{takano}, 
and more recently the
presence of superconductivity upon hole doping and under high pressure in 
the ladder compound
${ Sr_{0.4} Ca_{13.6} Cu_{24} O_{41.84} }$ has 
been reported~\cite{akimitsu}. Ladders appear to be near ideal systems 
where theoretical many-body calculations can be carried out with such
accuracy that their predictions 
can be contrasted directly with experiments. In the context of ladders
recently some
challenging results have  been experimentally
observed replacing $Cu$
by $Zn$ (a process which usually is considered equivalent to
 adding vacancies to the compound under investigation). In this situation
the spin gap of the pure system
 was observed to decrease very rapidly with $Zn$-doping, and,
even more surprisingly, antiferromagnetic order was found to be
 stabilized by this
procedure~\cite{azuma}. Such ordering effect caused by the apparently disordering 
addition of vacancies is puzzling.

In parallel with the recent excitement in the context of ladder systems,
another gapped low-dimensional compound has received much attention. 
It has been shown that $CuGeO_3$ has a spin-Peierls transition at $14K$
and at lower temperatures the compound is dimerized, with a finite spin
gap. While the addition of mobile carriers to this compound has not been
achieved, at least static
vacancies can be introduced by the same procedure as
in ladders i.e. replacing $Cu$ by $Zn$. It is a
remarkable experimental result that 
$Cu_{1-x} Zn_x Ge O_3$ was found to have 
a very similar behavior as $Zn$-doped ladders
(up to an overall energy scale) i.e. 
the original gap is rapidly suppressed and antiferromagnetic order
is stabilized after a small percentage of $Zn$ is introduced in the
system~\cite{hase1}. 

There are other compounds which present an intrinsic
structural dimerization  not caused by the coupling of spins and
phonons.  $Cu(NO_3)_2\cdot2.5H_2O$~\cite{bonner} and $CuWO_4$~\cite{lake},
provide examples of this behavior. More recently it has been shown that
$(VO)_2P_2O_7$, originally considered a two-leg spin ladder along the
$a$ direction, is actually better described by alternating spin chains
runing along the $b$ direction~\cite{garrett}. In addition, the compound
$Sr_{14}Cu_{24}O_{41}$ mentioned above
contains both two-leg
ladders and structurally modulated $CuO_2$ chains~\cite{hiroi}.

These experimental results have generated a large body of theoretical
work. Using a variety of many-body techniques and models, the issue of
the rapid collapse of the spin-gap upon the addition of
 vacancies in both ladders
and dimerized chains have been 
explained by the appearance of localized spin 1/2 states near the doped
vacancies~\cite{recent1d,recentladder,fukuyama}. For a very $Zn$-dilute 
system the average distance between these
states is large and, thus, the spins are weakly interacting.
Calculations
have shown that these localized states form a low energy band in the
spectrum which appear inside the original spin 
gap~\cite{recent1d,recentladder}. Predictions for
inelastic
neutron scattering (INS) experiment revealed that  for doped samples
while a large weigth should still exist
near a frequency $\omega$ corresponding to the original gap, 
extra weight growing proportional to
 $x$ should also be observed inside the gap~\cite{recent1d}. 
The results of recent INS experiments on doped $CuGeO_3$ are 
compatible with this prediction~\cite{azuma2}. 
Also Raman scattering results for in-chain and off-chain substitutions
in $CuGeO_3$ has been interpreted as signaling the presence of
 low energy excitations
inside the original gap, in agreement with the  theoretical 
studies~\cite{lemmens}. In addition, very recent
EPR experiments carried out for 
$CuGeO_3$ with up to $5\%$ $Zn$ doping have  revealed structure that
is also compatible with the presence of localized 
spins induced by vacancies~\cite{hassan}.
It is interesting to note that while all 
these  experimental techniques have enough resolution in energy to
separate features that smoothly evolve from the pure sample from those
that are induced by doping,
measurements of the static susceptibility may simply suggest 
the collapse of the gap
rapidly with $Zn$-doping since this quantity at low temperature is
dominated  by the low energy states. This was precisely the conclusion of the 
first papers on $Zn$ doped ladders and dimerized chains~\cite{hase1,azuma}.
Then, the experimental results
currently
available appear compatible among themselves and the presence of in-gap
states is very likely.
Thus, currently a consistent
scenario exists that could explain the early reports of an
apparent drastic reduction of
the spin-gap upon the addition of vacancies. This scenario relies on
spin 1/2 states induced near vacancies. Note that 
these spins will start interacting
as the vacancy concentration grows, thus they can be considered as ``free''
only at very small doping.

On the other hand, the 
appearance of enlarged antiferromagnetic (AF) 
correlations upon $Zn$ doping has been
more challenging to explain. Numerical calculations applied to dimerized
systems and ladders have indeed found that AF enhancement is a property
of the spin models used to describe their 
behavior~\cite{recent1d,recentladder}, but an intuitive explanation
is difficult to find. It is interesting to note that in other contexts
similar phenomena have been observed
before. For instance, staggered spin correlations are
amplified near vacancies in $S=1$ Heisenberg chains~\cite{sorensen},
presumably caused by the presence of localized $S=1/2$ states in its
vicinity. The $S=1/2$ Heisenberg model also has a staggered spin structure
which is enhanced near the edges of open chains, according to boundary
conformal field theory and Monte Carlo simulations~\cite{eggert}.
Finally, in the two dimensional (2D) Heisenberg model  studies
using a $4 \times 4$ lattice and spin-wave techniques detected the
presence of staggered order
 enhancement in the immediate vicinity of a vacancy~\cite{bulut}.

All these results have been studied mostly independently 
of each other in the literature. However, recently a simple
explanation for the AF enhancement has been
 proposed~\cite{prl} using the so-called ``pruned'' 
Resonating-Valence-Bond (RVB) picture~\cite{anderson}
which can be summarized as follows: at short distances the physics of
spin chains is dominated by spin singlet formation. For an undoped
system a given spin spends roughly half its time forming a singlet with
spin partners at both its right and left. However, if an impurity is located
next to the spin under consideration only one direction is now available
and this spin attempts to form singlets with its only 
 neighbor more strongly
than for the undoped case. Numerical and variational
results support this picture~\cite{prl}.

The purpose of the present  paper is twofold: (i) first, the ideas
introduced in the previous publication
Ref.~\cite{prl} by some of the present authors
are here expanded and the computational results are
discussed in detail. Some new results for
spin correlations are provided, and 
the generality of
the phenomenon of AF enhancement upon $Zn$-doping is emphasized.
 (ii)
Second, the study of 
experimentally observable consequences of the AF enhancement, specially
for NMR spectra, is discussed in this paper for a variety of models
with special emphasis on dimerized chains.
It is concluded that the anomalous broadening of the main
NMR signal as the temperature is reduced as predicted by
Eggert and Affleck~\cite{eggert} for compounds represented by the unfrustrated
undimerized $S=1/2$ Heisenberg  model will appear also
in other compounds with only small modifications. The techniques mainly
used
throughout the paper are the Lanczos~\cite{review} and Density Matrix
Renormalization Group (DMRG)~\cite{white} methods, but some results
at finite temperature
using the world-line Monte Carlo technique will also be presented. 
The physics discussed here is 
mostly caused by short distance effects, and thus finite size
problems are not expected to affect severely the results and conclusions
presented here.

\section{ Equal-Time Spin-Spin Correlations}

In this section numerical results are presented which illustrate
 the existence of
enhanced staggered spin-spin correlations near vacancies for a variety
of spin models and cluster geometries. These results suggest
 that the effect is quite
general and caused mainly by short distance physics, which makes  it  
 independent of the subtle properties of the ground state at  large
distances (i.e. gapped
vs gapless spin spectrum).

\subsection{$S=1/2$ Heisenberg Model With and Without Frustration}

The enhancement of staggered spin correlations upon the introduction of vacancies
appears clearly in the case of the
$S=1/2$ Heisenberg model defined on a 1D lattice,
as previously discussed in the literature~\cite{eggert}.
Fig.1 shows results obtained using the DMRG
technique~\cite{white} on a chain of 128 sites measuring
the staggered spin-spin
correlation $C_S(i,j) = \langle { {{\bf S}_i}\cdot{{\bf S}_j}} \rangle
(-1)^{i+j}$, where $\langle .. \rangle$ denotes the expectation value in
the ground state and the rest of the
notation is standard. The largest change between
the correlations at the center of the chain (bulk)
compared against those near  
 the edge (which simulates a $Zn$-impurity)
occurs at distance 1, i.e. the correlation among the first two spins next
to the vacancy is about $48\%$ larger than for a couple of
nearest-neighbor  spins located far from that vacancy.
The value obtained here for $C_S(1,2) = 0.652 $ is in excellent
agreement with recent Bethe Ansatz calculations on open spin chains
\cite{frahm} that reported $C_S(1,2) = 0.6515$.
The enhancement can also 
\begin{figure}[tbhp]
\centerline{\psfig{figure=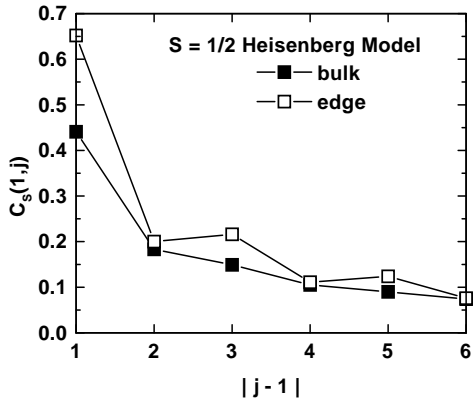,height=10cm,angle=0}}
\vspace{-4cm}
\caption[]{$C_S(1,j)$ for the spin 1/2 Heisenberg model without frustration
or dimerization (using the DMRG
method  on a chain with 128 sites). The open (full) squares are results
obtained locating site ``1'' next to the edge (at the center of the chain).}
\label{fig1}
\end{figure}
\noindent
be observed at larger distances specially when $|i-j|$ is odd
(note the ``zig-zag'' pattern of Fig.1 for ``edge'' results),
and it slowly decays with distance. If the results are  transformed to
momentum space, the real space enhancement translates into an increase
of the spin structure factor $S(q=\pi)$ (as remarked in Ref.\cite{eggert}).

Adding frustration in the form of a coupling $J_2 > 0$ 
between spins located
at a distance of two lattice spacings does not change qualitatively the effect
observed in Fig.1. The Hamiltonian is now
$$
H_{J_1 - J_2} = 
J_1 \sum_{i} { {{\bf S}_{ i}}\cdot{{\bf S}_{ i+1}} } +
J_2 \sum_{i} { {{\bf S}_{ i}}\cdot{{\bf S}_{ i+2}} },
\eqno(1)
$$
\noindent in the standard notation.
Fig.2 contains results for the case 
$J_2/J_1 = 0.3$ which corresponds to a coupling
where the ground state has a small but finite
spin gap in the spectrum. The spin correlations obtained with
both the Lanczos and DMRG
techniques provide very similar results.
The main effect of frustration is to reduce
the amplitude of these spin correlations at intermediate and large
distances (and it suppresses
 the zig-zag pattern of the unfrustrated case),
but at least at distance 1 the enhancement is still clearly present.
%Frustration seems to reduce the region where AF is favored upon Zn
%doping. 
\begin{figure}[h]
\centerline{\psfig{figure=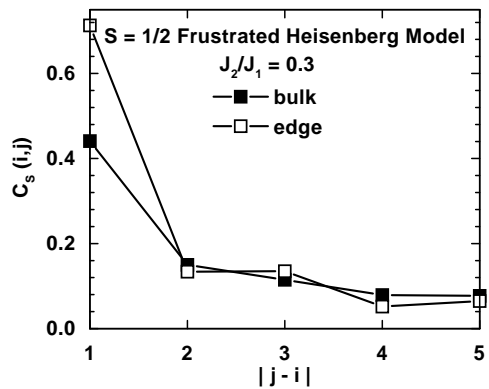,height=10cm,angle=0}}
\vspace{-4.1cm}
\caption{$C_S(i,j)$ for the $J_1 - J_2$ Heisenberg model at $J_2/J_1= 0.3$
(using the Lanczos  method  on a chain with 14 sites). 
The open squares are results
obtained locating site ``i'' next to the edge using open boundary
conditions (OBC).
The full squares denote results using periodic boundary conditions (PBC).}
\end{figure}

\subsection{Dimerized Chains}

The recent considerable effort devoted to the experimental and theoretical
 study of dimerized quasi-one dimensional
compounds motivated us to
investigate the enhancement of AF near vacancies for a simple dimerized
$J_1 - J_2 - \delta$ model defined as
$$
H = 
J_1 \sum_{i} (1 - (-1)^i \delta) { {{\bf S}_{ i}}\cdot{{\bf S}_{ i+1}} } +
J_2 \sum_{i} { {{\bf S}_{ i}}\cdot{{\bf S}_{ i+2}} },
\eqno(2)
$$
\noindent where the dimensionless parameter $\delta$ 
creates ``strong'' and ``weak''
links for the nearest-neighbor spin exchange. The rest of the
notation is standard. The presence of a finite
$J_2$ is added for completeness since some dimerized compounds seem to
have a sizable next-nearest-neighbors spin-spin
interaction~\cite{riera1}.
Note that the use of Hamiltonian Eq.(2) to study the influence of 
$Zn$-doping assumes that the pattern of strong and weak links does not
change upon doping. While this assumption is reasonable for 
the many structurally dimerized systems described in the Introduction,
it is questionable for spin-Peierls systems where phonons are dynamical
variables that could alter the dimerization pattern. This issue will be
discussed later in the paper.
The calculations using Eq.(2)
are very similar to those for the undimerized spin models of the
previous section. The Lanczos results corresponding to couplings expected
to be realistic such as  $\delta = 0.03$ and $J_2/J_1 = 0.2$
are shown in Fig.3 (DMRG results are not shown since they
are very  similar). 
Once again, a large enhancement of correlations is
observed at short distances, and here the zig-zag pattern of the gapless case
is still clearly visible. It is interesting to observe that at distance
1 (i.e. studying $C_S(i,i+1)$) 
 the enhancement obtained comparing  results at the edge and in the bulk
is larger than in the undimerized case.
To understand this effect consider three consecutive
spins labelled as ``0'', ``1'', and ``2'' on the chain, and assume that
the link between ``0'' and ``1'' is strong (and, then, that
the link ``1'' with ``2'' is weak). Without vacancies in the
neighborhood
the correlation between spins ``1'' and ``2'' is small since they share
a weak link. However, now assume that the spin ``0'' is removed (a
vacancy is added). In this case spin ``1'' has lost its
partner to form singlets with, and, as a consequence,
a free $S=1/2$ near the end is created (as
discussed recently in Ref.~\cite{recent1d}). This spin strongly attempts
to form a singlet with the only nearest neighbor spin available which is
``2'', 
enhancing substantially their correlation.
If the first pair of spins next to the vacancy 
\begin{figure}[tbhp]
\centerline{\psfig{figure=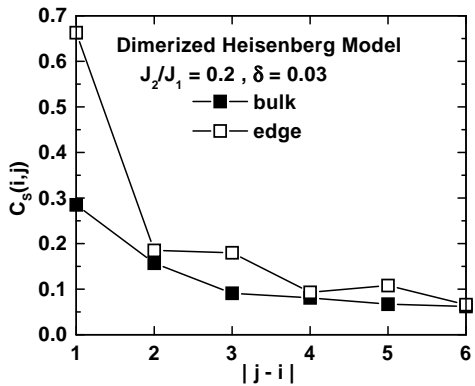,height=10cm,angle=0}}
\vspace{-4cm}
\caption{$C_S(i,j)$ for the 
$J_1 - J_2 - \delta$ Heisenberg model at $J_2/J_1= 0.2$
and $\delta = 0.03$
(using the Lanczos method  on a chain with 14 sites). 
The open squares are results
obtained locating site ``i'' next to the edge using OBC.
The full squares denote results using PBC.}
\end{figure} 
\noindent
were linked by a strong
bond rather than weak, then the enhancement would not be as 
dramatic. In other words, the
free spin induced by $Zn$-doping on a dimerized chain is located either at the
left or right of the vacancy, and the place where the AF enhancement
occurs is correlated with the location of that free spin.

\subsection{Ladders}

The existence of a vacancy-induced enhancement of antiferromagnetic
spin correlations is not restricted to chains. As shown below
the effect also appears on ladders and planes. As remarked
in the introduction this is of particular importance
since recent experimental results showed that
upon $Zn$-doping ladder compounds develop strong
AF correlations at low temperatures~\cite{azuma}.
Consider a $S=1/2$ Heisenberg model defined on a
2-leg ladder. Fig.4a shows $C_S(i,j)$ for a $2 \times 32$
cluster with DMRG results obtained
both along the leg where the impurity is located
and also along the opposite leg. The result shows that $C_S(i,j)$ is
specially enhanced for ``same leg'' correlations. This 
enhancement at distance 1, i.e. considering 
$C_S(i,i+1)$, is about $22\%$ which is
smaller than for a spin chain.

Fig.4b contains results for the case of a ladder where the
rung and leg couplings ($J_{\perp}$ and $J$, respectively)
 are selected such that the chains are not as
strongly coupled with each other as for the case of an
 isotropic ladder. No vacancy is introduced here but
simply the correlations are measured with open and 
periodic boundary conditions, 
starting in the former from the edge of the ladder.
In this situation the spin
correlations resemble qualitatively the results found before for
the $S=1/2$ Heisenberg chain including the zig-zag pattern
in the enhancement. 
\begin{figure}[tbh]
\centerline{\psfig{figure=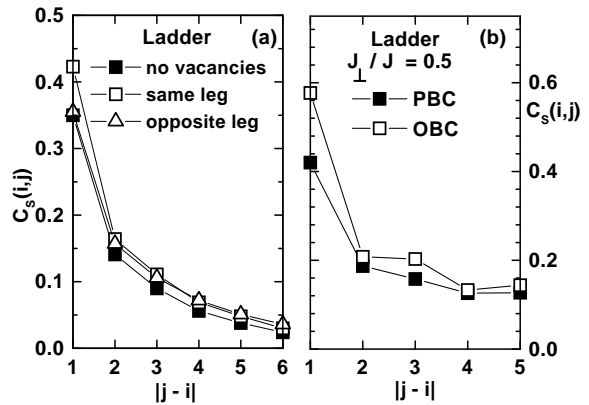,height=10cm,angle=0}}
\vspace{-4cm}
\caption{(a) $C_S(i,j)$ for a two-leg ladder calculated with DMRG on a $2
\times 32$ cluster. Open squares (triangles) denote spin
correlations along the same (opposite) leg where the vacancy is located,
with a starting site $i$ next to the vacancy (which itself is at the
center of the cluster).
Full squares are results
without vacancies obtained in the bulk of the cluster; (b) $C_S(i,j)$
for a two-leg ladder calculated with the Lanczos method on a $2
\times 10$ cluster at coupling
$J_{\perp}/J = 0.5$. Full squares are results with PBC. 
Open squares are results with OBC and
correlations measured from the edge of the cluster.}
\end{figure}
\noindent
Studying a variety of ratios $J_{\perp}/J$ 
it was observed that the correlations at short distance smoothly
interpolate between those of decoupled chains ($J_{\perp}/J = 0$) 
and the isotropic limit
($J_{\perp}/J = 1$). This suggests that the effect is general
and does not depend on the gapped vs gapless character of the ground 
state, in agreement with
the conclusions from  the study of the  $J_1-J_2$ chain
(only the size of the disturbance is a function of the long distance
properties of the Hamiltonian under investigation).

\subsection{Two Dimensions}

The existence of enhanced spin-spin correlations
due to the presence of a vacancy
is not restricted to one dimensional systems. A
similar effect occurs also in the two dimensional (2D)
Heisenberg model, as previously reported in Ref.\cite{bulut} for
the unfrustrated case.
Fig.5a-b shows the spin-spin correlations
calculated on a tilted $\sqrt{26} \times \sqrt{26}$ 
Heisenberg cluster with and without a vacancy, and including
a frustrating coupling across the plaquette diagonals of
strength $J_2$ (with $J_1$ being the nearest-neighbor coupling). 
Results for a $\sqrt{20} \times \sqrt{20}$ cluster are 
qualitatively similar. The first
site $i$ in the correlation is next to the vacancy, and
$j$ runs over neighbors at distance $1$, $\sqrt{2}$, $2$, and $\sqrt{5}$
away from $i$. Fig.5a-b show
that the enhancement effect is mostly restricted to
distance 1. In the unfrustrated case $J_2/J_1 = 0.0$, the change
of the correlation for the special case $|i-j|=1$, denoted by $\Delta_1$,
with and without the impurity is about $11\%$. 
\begin{figure}[tbhp]
\centerline{\psfig{figure=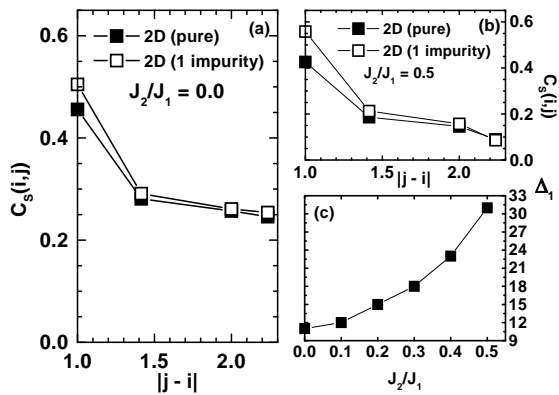,height=10cm,angle=0}}
\vspace{-4cm}
\caption{(a) $C_S(i,j)$ on a $\protect\sqrt{26} \times \protect\sqrt{26}$
tilted square cluster calculated with exact diagonalization. 
Open (full) squares denote results with (without) a vacancy. $i$ is
the site next to the vacancy if present. Correlations at distances
$1,\protect\sqrt{2},2$ and $\protect\sqrt{5}$ away from site $i$ are shown;
% and the measurement is along the axis that
%contains the vacancy. 
(b) Same as (a) but introducing frustration using the coupling
$J_2/J_1=0.5$;
(c) Relative enhancement $\Delta_1$ (in $\%$) of the spin-spin correlation
 at distance of one lattice spacing (between a site next to the
vacancy and its neighbor in the direction away from the vacancy)
vs. $J_2/J_1$.}
% and the measurement is along the axis that
% contains the vacancy. 
\end{figure}
%\noindent

On the other hand,
at $J_2/J_1 = 0.5$ i.e. in a
situation where the frustration
is strong enough to melt the antiferromagnetic
order (but not strong enough to make the system collinear~\cite{j1j2})
$\Delta_1$ is about $31 \%$ (Fig.5c shows $\Delta_1$ vs $J_2/J_1$ for
other couplings).
It is clear that the enhancement occurs more easily when the
tendency to form an antiferromagnetic ground state 
is suppressed i.e. when 
the disordering effects are strong. 
However, once again
the existence of the  effect itself does not seem much
related with the long distance ground state properties of the system. 
It is likely that the enhancement is
caused by the behavior at $short$ distance 
which likely is similar to frustrated and unfrustrated Heisenberg models.

\subsection{1D $S=1$ Heisenberg Model } 

The results described in this section are not restricted to $S=1/2$
systems but the AF enhancement also appears in models with a higher
spin, as shown below.
%However, the increase in the correlations 
%seems maximized for $S=1/2$. 
This can be understood intuitively
recalling that at zero temperature a classical spin system
$(S \rightarrow \infty)$ is antiferromagnetically ordered even 
on a chain with open boundary conditions (OBC). Then, while in the $S=1/2$
case there
is a clear AF enhancement, for $S=\infty$ the effect is not
present in the ground state. As a consequence, it is
reasonable to expect that for
spins such as 1
the  spin-spin correlation enhancement should roughly interpolate
between $S=1/2$ and $\infty$.
%, and as a consequence it may be
% smaller than for the $S=1/2$ spin systems
%discussed in previous subsections. 
Fig.6a shows
$C_S(i,j)$ for a 1D $S=1$ Heisenberg model. 
This clear AF enhancement observed in Fig.6
was extensively discussed in previous literature and it was
attributed to $S=1/2$ states localized near 
\begin{figure}[tbhp]
\centerline{\psfig{figure=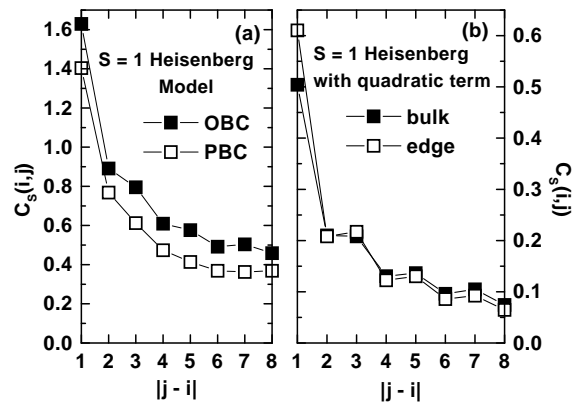,height=10cm,angle=0}}
\vspace{-4cm}
\caption{$C_S(i,j)$ for the 1D $S=1$ Heisenberg model
obtained with the DMRG technique on a chain with 32 sites and
keeping $m=27$ states in the iterations. The notation is as in
previous figures. (a) denotes results for the pure Heisenberg
model, while (b) has results adding a
$- J \sum_{  i } 
({ {{\bf S}_{ i}}\cdot{{\bf S}_{ i+1}} } )^2$ term to the Hamiltonian
(inducing a gapless ground state).}
\end{figure}
\noindent
the edges~\cite{sorensen}.
Although 
the relative enhancement at distance 1 is smaller than for the case
$S=1/2$, it is remarkable that the effect extends
over a surprisingly large distance. Fig.6b shows results adding
a term 
$- J \sum_{i } 
({ {{\bf S}_{ i}}\cdot{{\bf S}_{ i+1}} } )^2$ to the Heisenberg model.
This particular addition  renders the ground state
gapless~\cite{deisz}.
The results for the spin correlations show that in this case 
there is an enhancement in the correlation which
has a staggered pattern very similar to the one observed for the 
$S=1/2$ unfrustrated model (Fig.1). This example shows that 
qualitatively the
effect discussed here will likely appear for any finite $S$ spin chain
 and for
both gapped and gapless models (although its actual intensity 
depends on the details of the  ground state under
investigation).

\section{``Pruned'' RVB Basis}

The results shown in the previous section suggest that a universal
mechanism may be at work since a variety of models and cluster geometries
present the same phenomenon of  enhanced antiferromagnetism
near vacancies.
In this section a tentative
explanation for this effect
is given in terms of the Resonating-Valence-Bond states
in its short-range version i.e. with states obtained from a covering of
the lattice with short dimers~\cite{kivelson}.
The use
of these states to explain the $short$ distance phenomenon studied here
can be justified even if globally such short-range RVB states do not
properly represent the ground state properties of the model under
consideration. A typical example is given by the 2D
antiferromagnetic Heisenberg model: in this case it is well-known that 
at zero temperature the model
is gapless, with long-range order, and that it cannot be represented by a
combination of short-range RVB states since they have a 
spin gap. However, numerical studies have
shown~\cite{overlap} 
that the exact ground states of $finite$ clusters actually have a
large
overlap with the short-range RVB state for the same cluster. This
suggests that at short distances even an asymptotically
 ordered system such as a 
2D Heisenberg model can be approximated by a superposition of spin singlets.
Of course, to obtain the proper long-range behavior in the bulk
spin singlets linking sites at large distances are needed.

Based on this discussion the following scenario is proposed~\cite{prl}.
Consider Fig.7a,b where a couple of  configurations with short
 spin singlets are represented. These spin arrangements are expected to be
important for a variety of spin models in 1D, and in particular it
is the exact ground state for the frustrated Heisenberg model with
$J_2/J_1=0.5$~\cite{mg}. 
Consider now a vacancy introduced at the site labeled as
``0'' and let us study what happens with the configuration shown in
Fig.7b: in this
case a free spin is generated next to the vacancy (schematically shown
in Fig.7c). This free spin
will tend to couple strongly with its neighbor and, thus, the
configuration Fig.7d will have a high probability as the system evolves
in time. Eventually, the free spin can move some
 distance away from the vacancy following
the same procedure and configuration 
Fig.7e is reached. For the case of the $S=1/2$ Heisenberg
model this ``free''
spin is not bounded to the
vacancy. This is a manifestation of the well-known effect of spin-charge separation
in the case of the $t-J$ model in one dimension for the special case of
static holes (in the rest of the text when
 ``spin-charge separation'' is invoked it must be
understood in this framework).For other models such as the 2-leg ladders
or dimerized systems the free spin can be localized
\begin{figure}[tbhp]
\centerline{\psfig{figure=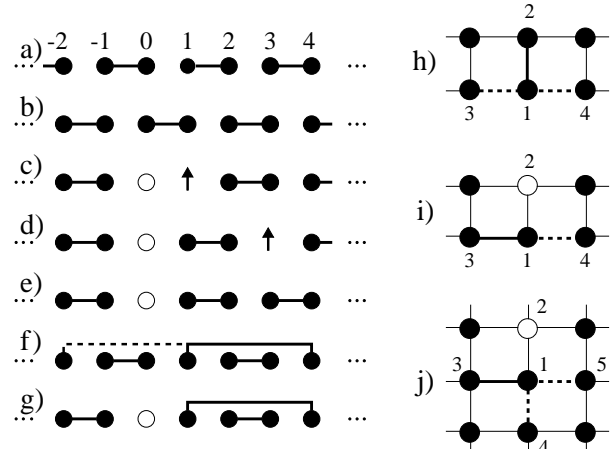,height=6cm,angle=0}}
\vspace{0cm}
\caption{(a)-(g) Examples of spin singlets relevant for the ``pruned'' RVB
scenario proposed as an explanation for the numerical results;
(h)-(j) Spin singlets relevant for ladders and planes. See text for details.
(taken from Ref.~\protect\cite{prl}).}
\end{figure}
\noindent
in a finite region near the
vacancy~\cite{sandvik,recent1d}. But as
long as the size of this spin-vacancy bound state
 is more than three or four lattice spacings the 
reasoning that led to Fig.7a-e 
can be applied.
The conclusion is that the spin next to the vacancy 
(spin ``1'') forms
a singlet with only one partner after spin ``0'' is removed
simply because there is only one partner available, while
without vacancy doping
the spin  at ``1'' would spend equal
amounts of time coupled with spin partners at both
its right and left. This effect
naturally causes an enhancement of the correlation 
$\langle { {{\bf S}_1}\cdot{{\bf S}_2} } \rangle$ due to the
introduction of the vacancy. $Zn$ doping effectively $prunes$
the set of possible RVB
spin configurations from 2 down to 1. Spin ``1'' no longer resonates but its
partner is fixed by geometry.

The same reasoning applies if singlets longer than one lattice spacing
are considered. Fig.7f illustrates the situation where the spin ``1''
is coupled with ``4'' through a singlet of length 3. In this case by
symmetry arguments
in the wave function a singlet between ``1''
and ``-2'' will exist carrying 
equal weigth. However, once again by
cutting the chain introducing a vacancy at ``0''
only one possibility remains as in the case of the short spin singlets,
and the correlation between ``1'' and ``4'' is enhanced (Fig.7g). The same
reasoning can be repeated for longer singlets, although their importance
decays with distance and eventually results close to those obtained 
far  from the edge should be obtained when
$C_S(1,j)$ is calculated.

The argument presented here 
leads to the prediction that the AF enhancement should be
maximized in 1D. The idea is illustrated in Fig.7h for the case of a
2-leg ladder geometry: if a vacancy is introduced at a given site, the spin
next to it along the same rung now has only 2 possible spin
partners  to form singlets
while before it had 3. Thus, the spin correlation between the spin
under study and its 2 neighbors will be
enhanced~\cite{comment2}. Repeating the same argument for
a 2D system, 
the ratio of partners before and after the vacancy is 4
and 3, respectively. In general for an arbitrary dimension $D$, the
ratio
is $(2 \times D)/ (2\times D -1)$ which tends to 1 as $D$ grows. 
Then, one dimensional systems are the best candidates to present
a substantial enhancement of correlations due to vacancies. The
previous numerical results in 2D clusters showing a spin enhancement
smaller than in 1D supports this reasoning. 
%It is unlikely
%that 3D models will have a noticeable enhanced correlation in this context.

These ideas can be further tested using RVB-like 
variational wave functions on
small clusters. Using the notation $(ij)$ to denote a singlet between
the spins at sites $i$ and $j$, and $[i,j]_\sigma$ for a triplet between
the same two spins with total projection $\sigma=1,0,-1$, the proposed
state is
$$
|RVB \rangle = | (12)(34)(56)(78) \rangle +
\alpha (~|(14) (23) (56) (78) \rangle +
$$
\vspace{-1.2cm}

$$ ~~~~~~~~| (12) (36) (45) (78) \rangle
+ | (12) (34) (58) (67)  \rangle ~) +
$$
\vspace{-1.2cm}

$$
~~\sum_{\sigma,\sigma'} 
\beta_{\sigma,\sigma'} (~| [13]_{\sigma} [24]_{\sigma'} (56) (78)
\rangle + | (12) [35]_{\sigma} [46]_{\sigma'} (78) \rangle +
$$
\vspace{-1.2cm}

$$
~~~~~~~~~~| (12) (34) [57]_{\sigma} [68]_{\sigma'} \rangle ~~),
\eqno(3)
$$
\noindent where the set $\{ \sigma,\sigma' \}$ takes the values $(1,-1)$,
$(-1,1)$, and $(0,0)$, and $\beta_{1,-1} = \beta_{-1,1}$ by symmetry.
This state is more general than the discussion presented before since it also
contains spin triplets that are needed to quantitatively reproduce the
spin correlations for the clusters studied. The concept of ``pruned''
RVB basis can be easily extended to include spin triplets.

Results are shown in Fig.8. Short chains with OBC
are considered to simulate the effect of vacancies.
\begin{figure}[tbhp]
\centerline{\psfig{figure=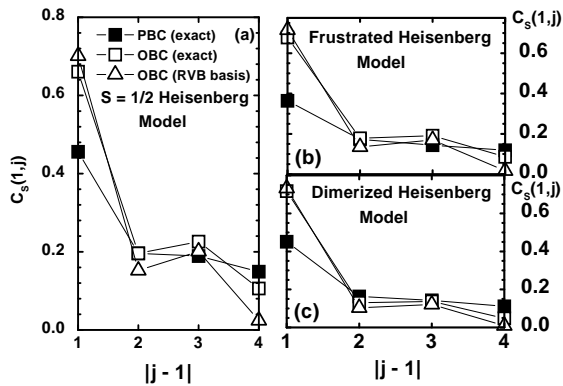,height=10cm,angle=0}}
\vspace{-4cm}
\caption{Staggered spin correlations calculated using 
RVB variational states (see text) on an 8 site chain with 
OBC, contrasted against exact results with both OBC and  PBC.
(a) corresponds to the $S=1/2$ Heisenberg model; (b) frustrated $J_1 -
J_2$ model with $J_2/J_1 = 0.3$; and (c) dimerized $J_1 - J_2 - \delta$
model with $J_2/J_1 = 0.2$ and $\delta = 0.03$.}
\end{figure}
\noindent
For the frustrated and unfrustrated
$S=1/2$ Heisenberg model, as well as the dimerized chain, the variational
results are very close to those obtained exactly, and both differ from 
the correlations for the same lattice size but using PBC
where there are no edge effects. 
The variational state gives remarkably accurate results at
 distance 1, and since it is dominated by the configuration
with the shortest spin singlets it naturally reproduces the zig-zag pattern of some
correlations at larger distances. 
Then, these results suggest that the
simple picture presented in this section is robust and the
antiferromagnetic
enhancement is caused by a pruning of the set of short-range RVB states
that describe the short distance behavior of the spin correlations
for a variety of spin models.

\section{Influence of Coulombic Interactions and Density}

The spin models analyzed thus far can be considered as the
strong Coulomb coupling limit of suitable chosen
electronic models. In particular, the $S=1/2$ Heisenberg
model is recovered from the one band Hubbard model
with nearest-neighbor hopping at large $U/t$. Then, it
is interesting to analyze the  enhancement of AF spin
correlations discussed in this paper when $U/t$ is varied, and also
as a function of the electronic
density $\langle n \rangle$ away from half-filling~\cite{ulmke}. Fig.9
shows results for the Hubbard model at $U/t=10$, $4$ and $0$,
and $\langle n \rangle = 1.0$ (half-filling)
as representative of the strong, intermediate, and weak
coupling regimes, respectively (note that the bandwidth is
$W=4t$ in 1D).
\begin{figure}[tbhp]
\centerline{\psfig{figure=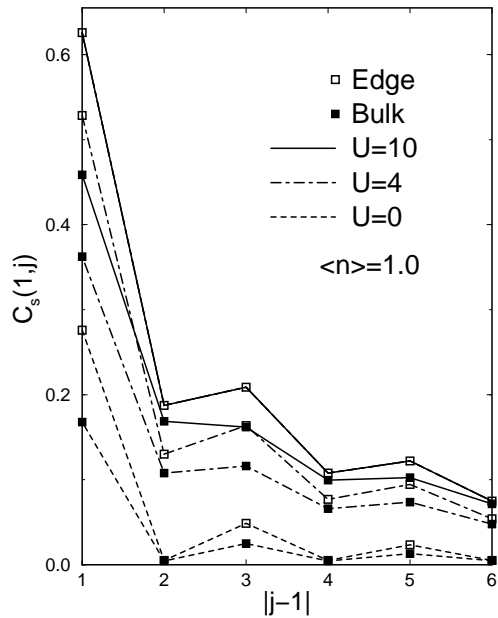,height=10cm,angle=0}}
\vspace{-0.7cm}
\caption{$C_S(1,j)$ for the 1D Hubbard model using the DMRG technique
on a 60 sites chain working at $\langle n \rangle =1$ and the couplings shown
($t=1$). The open
(full) squares are results obtained locating site ``1'' next to the edge (at
the center of the chain).}
\end{figure}
\noindent
As $U/t$ is reduced a smooth change in the spin correlations
is observed. These correlations decrease in absolute value
but the relative  enhancement remains strong at short distances.
The ``zig-zag'' pattern described in Sec.II is present even in the
non-interacting limit $U/t=0$.
% where it resembles
%Friedel-like oscillations. 
The example of the Hubbard model at weak coupling show us that the
interpretation of the effect based on the RVB language of Sec.III
may not be unique, and other explanations based on perturbative
approaches in the small $U/t$ limit
could lead to similar conclusions. This is reminiscent of the
presence of bulk 
strong antiferromagnetic correlations themselves which can be
understood as a ground state property of a spin localized Heisenberg
model, as well as arising from nesting properties of the half-filled
noninteracting Hubbard model. In other words, $U/t=0$ is a critical
point for the one band Hubbard model with nearest-neighbor hopping and
it already has strong indications of the existence of antiferromagnetic
correlations in the ground state, although they are not of long range
order. 
%It is natural that the enhancement of antiferromagnetism
%by vacancies discussed in this paper exists as long as there is some
%antiferromagnetism in the bulk   ground state of the model under 
%consideration.

Let us consider now the influence of density on the AF 
vacancy-induced enhancement.
It is reasonable to expect that the effect will be observed as long as
the bulk ground state spin correlations are dominantly staggered.
Since previous experience
with Hubbard-like models in 2D has shown that AF correlations are
rapidly suppressed by doping~\cite{review}, the same will likely occur with
the AF enhancement near $Zn$-ions. Results at densities away from
half-filling are presented in Fig.10 for $U/t=4$. 
\begin{figure}[tbhp]
\centerline{\psfig{figure=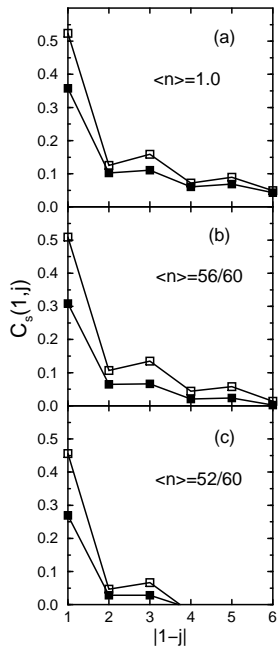,height=10cm,angle=0}}
\vspace{-0.3cm}
\caption{$C_S(1,j)$ for the 1D Hubbard model using the DMRG technique
on a 60 sites chain at $U/t=4.0$ and the densities shown. 
The open
(full) squares are results obtained locating site ``1'' next to the edge (at
the center of the chain).}
\end{figure}
\noindent
Indeed, the range of the 
AF correlations is certainly  suppressed as $\langle n \rangle$
deviates just a small percent from half-filling. 
However, at very
short distances such as 1,2, and 3 lattice spacings the effect remains
visible. Specially the tendency to form a robust spin singlet
between the 
two spins located the closest to the vacancy (or edge of an open chain)
seems to survive the rapid reduction of the AF tail in the spin correlations.
This result may have consequences for experimental investigations of
the phenomenon investigated in this paper. For instance, in the normal
state of underdoped cuprates, where it is believed that a AF
correlation length of about a couple of lattice spacings should
exist, $Zn$-doping
may induce localized moments and small regions with 
nonzero staggered correlations. However, 
more work is needed to fully clarify the
consequences of the vacancy-induced AF effect in cuprates
since a combination of mobile charges and static defects may complicate
the interpretation of experiments.

\section{Local Susceptibility}

After studying in detail the equal-time spin correlations,
in this section some experimental consequences of the
vacancy-induced enhanced antiferromagnetism are described.
The emphasis is given to NMR experiments where the measured spectra 
depends  on the local spin susceptibility. The discussion here
follows previous calculations
in the context of the $S=1/2$ Heisenberg chain where
the theoretical predictions for the NMR spectra~\cite{eggert}
have been recently confirmed experimentally~\cite{takigawa}.

\subsection{Intuitive Expected Behavior:}

The local susceptibility corresponding to site $i$ and temperature $T$
for any of the spin models used in Sec.II is defined  as
$$
\chi_i(T) = (1/T) \sum_j \langle  S^z_i S^z_j \rangle,
\eqno(4)
$$
\noindent in a standard notation.
Note that as $T \rightarrow 0$ and working on a finite
lattice, both the numerator and denominator vanish.
The reason is that for spin systems of the family
considered here the ground states are usually spin singlets which are
 separated
from the excited states by a finite gap (which may be intrinsic
to the model or caused by finite size effects). In such a situation 
$\sum_j S^z_j$ acting over the ground state produces a vanishing response.
Then, in order to study the local susceptibility using the DMRG or
Lanczos methods,
which are zero temperature algorithms applied to finite clusters, 
the effect of a finite
temperature needs to be simulated.  For this purpose 
the first excited state in the energy spectrum
having a finite spin (which
typically is spin one or one-half depending 
on whether the number of sites is even or odd)
 will be used in Eq.(4) to obtain the $\langle .. \rangle$ average. 
This state is the first that contributes appreciably to the
susceptibility when
temperatures or magnetic fields
of the order of the gap in the spectrum are considered, and here it will
be denoted 
by $| \phi_{S} \rangle$ where $S$ is 1 (1/2) for an
 even (odd) number of spins.
Then, the low temperature
local susceptibility used in the DMRG/Lanczos calculations below
is redefined as
$$
\chi_i \approx \sum_j \langle \phi_{S} | S^z_i S^z_j | \phi_{S} \rangle
= S~\langle \phi_{S} |
S^z_i | \phi_{S} \rangle,
\eqno(5)
$$
\noindent where in the last step the equality
$\sum_j S^z_j | \phi_{S}
\rangle = S | \phi_{S} \rangle $ was used. Comparing results using this
approach  with those of finite
temperature Monte Carlo simulations  presented later in the text,
 it will be  shown that this
definition captures not only qualitatively but even quantitative
the important aspects of the low
temperature behavior of the actual susceptibility $\chi_i(T)$
(for bulk  gapless models this good agreement is expected for 
finite systems only i.e. when
there is a small gap in the spectrum
caused by size effects).
Note also that the factor $1/T$ being just a multiplicative factor
will simply be dropped from the calculations below, and thus the
units in the reported local susceptibilities are arbitrary.

Let us discuss the expected qualitative behavior of $\chi_i$ as
defined in Eq.(5) considering as example the spin correlations
of a dimerized $J_1 - \delta$ model in the limit of a large enough
$\delta$ (i.e. working in a ground state with a robust spin gap)
such that the spins far from the chain edges
mainly form
tight spin singlets with their neighbor across the ``strong'' bonds
of the chain. Let us also assume  that the first link next to the
edge is ``weak''. In
this situation it is intuitively reasonable (and Fig.3 confirms it)
 that the first two spins near the end will attempt to form a singlet
at least part of the time. Fig.11a provides a snapshot of a typical
spin configuration where a mismatch between the bulk spin singlet
pattern and the spin singlet near the end can be observed. 
\begin{figure}[tbhp]
\centerline{\psfig{figure=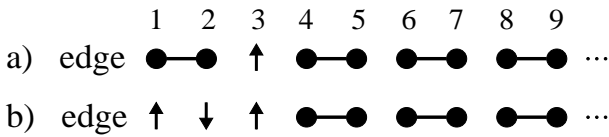,width=8cm,angle=270}}
\vspace{0.3cm}
\caption{(a-b) Two spin configurations relevant for the discussion on
the development of a maximum in the local susceptibility close to an
edge. See text for details.} 
\end{figure}
\noindent
At the
location of the defect (which is a standard domain wall) a quasi-free
spin 1/2 is located. It is clear that the local susceptibility for
Fig.11a will have the value 1/4 at site 3 where the free spin is 
(since $\langle S^z_3 S^z_{j \neq 3} \rangle = 0$), while at all other
sites it will cancel since $\langle S^z_i S^z_i \rangle 
+ \langle S^z_i S^z_{i+1} \rangle = 1/4 - 1/4 = 0$ (if $i$ and $i+1$ 
are the locations of the spins forming a strong singlet)
and $\langle S^z_i S^z_j \rangle = 0$ if $j \neq i+1$. 
Note, however, that the location of the free spin can certainly fluctuate
when the full quantum mechanical calculation is carried out.
In the case of the dimerized chain at hand, if the free spin has moved some
distance away
from the vacancy the spins in between form the ``wrong'' vacuum
i.e. they have the pattern of singlets on the weak links rather than
the strong ones. This argument clearly shows that the free spin is localized
near the chain end or vacancy, occupying
a finite region of extension
 $\lambda$. It is reasonable to expect that the
combination of configurations in the ground state of
 the localized spin (again, carrying out a full quantum calculation)
will produce a spin susceptibility nonzero for $i < \lambda$ and
approximately zero outside. Inside the bound state the spin configuration
will be scrambled or disordered
 by the movement of the spin 1/2 defect, while outside
it rigidly forms a dimerized pattern.
In addition, it is also likely that the spin 1/2
may behave in the interval $1 \leq i \leq \lambda$ as a particle in a
confining potential and, thus, its wave function will peak at a site
away from the extremes. In this situation, $\chi_i$ will have a $maximum$
at that particular site $i$ inside $i < \lambda$.
Finally, note that inside the bound state where the individual
spins do not follow
the same pattern as in the bulk, the antiferromagnetic nature of the
Hamiltonian will induce an $alternation$ in the sign of $\chi_i$.
As a toy model to understand this effect
consider Fig.11b where near the edge 3 spins are considered
to represent the $S=1/2$ bound state, and the rest is forming very strong
dimers. Solving exactly the 3 spin Heisenberg
problem in the $S^z_{total}=1/2$ 
subspace, the ground state has energy $-J$, 
the eigenvector is $| \psi_0 \rangle = \sqrt{2/3} [ | \uparrow \downarrow
\uparrow \rangle - (1/2) ( | \uparrow \uparrow \downarrow \rangle + 
| \downarrow \uparrow \uparrow \rangle ) ]$,
and the spin correlations are 
$\langle S^z_1 S^z_2 \rangle = \langle S^z_2 S^z_3 \rangle = 1/6$
and $\langle S^z_1 S^z_3 \rangle = 1/12$
(the ground state can also be written as a combination of the state with
a spin singlet between spins 1-2 and a spin up at 3, and the state with
a spin singlet between spins 2-3 and a spin up at 1).
The spin correlations produces susceptibilities
$\chi_1 = \chi_3 = 1/6$ and $\chi_2 = -1/12$, showing the expected
alternation. In other words, the spin at the center wants to point mostly down
due to the influence of the two neighbors which are mainly up, and then
the susceptibility has to be negative.
$|\chi_i|$ is here maximized at the sites where the chances
of having a spin up are the largest which are 1 and 3 (the 3 spin problem
is too small to show a maximum away from the edges as it occurs on larger
systems).  A peak
in $\chi_i$ away from $i=1$
is expected for cases where $\lambda$ is larger than 3 sites
since spins at the center near $i \sim \lambda/2$ will be influenced by
all the rest of the spins, while those near the end by only half of
those spins.

Consider now a gapless system as a limit of a spin gapped model. It is
intuitively plausible that as the spin gap of a given Hamiltonian is reduced,
for instance by properly
adjusting some couplings such as $\delta$ in the dimerized chain, 
the position of the maximum will
move away from the chain end. 
%In other words, the existence of a
%localized
%$S=1/2$ state near the edge is apparently correlated with the existence
%of a spin gap in the spectrum.
In the limit where the gap
vanishes, $\chi_i$ could  become a monotonously growing function of 
distance $i$. This argument makes the results of Eggert and 
Affleck~\cite{eggert} intuitively
reasonable, although certainly do not
replace the conformal field theory methods and Monte Carlo simulations
employed by those authors. This expected behavior for 
gapless systems is observed in the 
DMRG studies for the $S=1/2$ Heisenberg model.
%given below using the lowest energy state in the subspace
%of spin 1, as explained before in this Section.

The local susceptibility is a symmetric function under reflexions with
respect to the
center of the chain studied, and it has rapid sign oscillations.
Then, it can be separated into a
``uniform'' component $\chi_i^{u}$ 
and a ``staggered'' or ``alternating'' component using the definition
$\chi_i = \chi^u_i - (-1)^i \chi^a_i$. Both of these components are now
smooth functions of the site index. In practice the uniform part was
obtained below using $\chi^u_i \approx {{1}\over{2}} \chi_i + {{1}\over{4}} 
( \chi_{i+1} + \chi_{i-1} )$ and $\chi^a_i = -(-1)^i ( \chi_i - \chi^u_i
)$, where the former equation provides an approximate but accurate estimation
of the uniform part for smooth functions of the site position.
With these definitions the staggered component becomes an antisymmetric (symmetric)
function of $i$ for an even (odd) total number of sites (the proof is by simple
explicit construction).
In addition, it can be shown that as the chain size grows
the sum $\sum_i \chi_i^{u}$ 
converges to the total spin in the $z$-direction 
of the state $| \phi_{S} \rangle$ considered in Eq.(5)
(i.e. 1 (1/2) for an even (odd) number of sites).
Then, $\chi_i^{u} \sim
1/N$ at the center of the chain if a gapless model is considered
(i.e. the spin of the state is approximately uniformly
distributed on the chain). For systems
such as a dimerized lattice
where a localized spin 1/2 appears near the ends of the chain
and there is a spin gap,
$\chi_i^{u}$ should vanish in the bulk of the chain and remain
nonzero only at the ends. In the next subsections the qualitative expected
behavior of $\chi_i$ discussed here will be tested using numerical techniques.

\subsection{$\chi_i$ in the $S=1/2$ Heisenberg chain}

In Fig.12a the results for $\chi_i$ obtained with the DMRG method
using a chain with 80 sites are shown (remember that the DMRG technique
is applied with OBC i.e. it already simulates the presence of 
a couple of vacancies in the problem). 
The uniform and staggered
components satisfy the expected behavior described before.
As anticipated, $\chi_i^{a}$ 
initially grows with the distance $i$ from the end of the chain
as it should occur in the bulk~\cite{eggert}. 
However, Fig.12a and the previous
discussion show that symmetry
considerations  force $\chi_i^{a}$ to vanish in the middle of the chain,
changing sign between the left and right parts of the cluster,
which is a finite size effect.
The maximum in
$\chi_i^{a}$ is located at a position slightly smaller than
 $1/4$ of the chain length. Roughly, it is expected that
the region between the first site $i=1$ and the site where the
maximum is obtained is representative of the bulk.
Fig.12b shows again the two components of $\chi_i$ but now for a
chain with an odd number of sites. Once again in agreement with the
previous discussion now $\chi_i^{a}$ is symmetric. 
\begin{figure}[tbhp]
\centerline{\psfig{figure=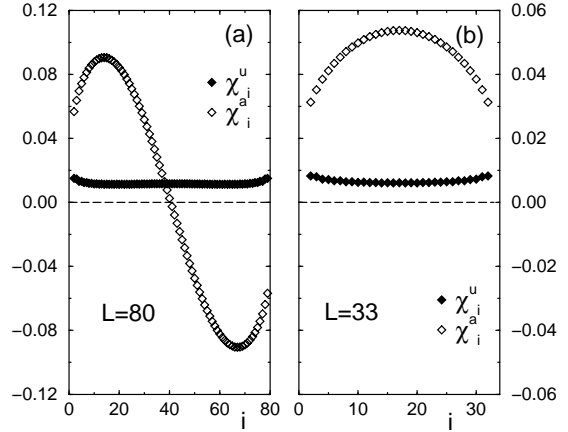,height=6cm,angle=270}}
\vspace{0.3cm}
\caption{Uniform and staggered components of the local susceptibility
$\chi_i$ obtained with the DMRG method applied to the $S=1/2$
Heisenberg model on a chain with $L$ sites, using $m$ states in
the iterations and studying the subspace with a total spin
in the $z$-direction equal to 1.
(a) corresponds to $L=80$ and $m=16$, and (b) to $L=33$ and $m=32$.}
\end{figure}
\noindent
It also
grows away from the chain ends as for an even number of sites, but its maximum
is reached in the middle of the chain. Then,
the finite size effects are less severe in a chain with an odd number of
sites.
Fig.13 illustrates the expected behavior in the bulk
analyzing short distances near the chain end using systems of length
$N=62,80$ and $160$.
\begin{figure}[tbhp]
\centerline{\psfig{figure=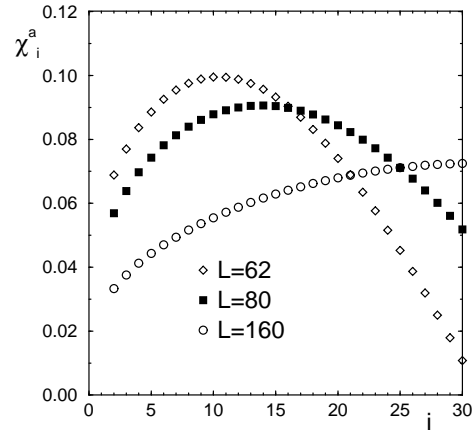,height=6cm,angle=270}}
\vspace{0.4cm}
\caption{Staggered component of $\chi_i$ for the $S=1/2$ Heisenberg model
on a chain, obtained with DMRG keeping $m=16$ states. Results for several
lengths are given.}
\end{figure}
\noindent
The position of the maximum moves away from the chain end.
For a long enough chain a monotonous increase of $\chi_i^{a}$ is recovered
in agreement with the calculations of Eggert and Affleck~\cite{eggert}.

\subsection{$\chi_i$ in the frustrated $S=1/2$ Heisenberg chain}

Fig.14a,b contain results similar to those of Fig.12a,b but now introducing
frustration in the Hamiltonian using a Heisenberg interaction at
distance
of two lattice spacings regulated by the coupling $J_2$. For
the particular case $J_2/J_1 = 0.5$ the ground state is known to 
consist of spin dimers at alternating links, and it has a spin-gap
(on a finite chain with PBC, it is doubly degenerate)~\cite{mg}.
In spite of such clear  differences between the ground states at
 $J_2/J_1 = 0$ and $0.5$, the
local susceptibility behaves very similarly as shown in the figure.
The uniform component does not show any sign of having a localized spin
state near the ends 
 (actually the spin distribution has its minimum at the edges). Then,
the spin  $J_1 - J_2$ model has a finite spin gap and yet no localized
spin 1/2 states near the edges.
A very similar behavior was
observed in the case $J_2/J_1 = 0.4$ (not shown), and, thus, the results
of Fig.14a,b can be considered as representative of the
spin gapped regime of the frustrated Heisenberg chain. 
An enhanced staggered susceptibility does 
not only exist in the unfrustrated Heisenberg model but also in a much
broader family of models that include frustration in the Hamiltonian. 
The key common
ingredient to obtain quantitatively a growing $\chi_i$ away from the
edges seems the
absence of localized spins at the ends. 
%In this situation (spin-charge
%separation) the qualitative results found by Eggert and Affleck apply.
Thus it is here concluded that the
 phenomenon of a growing staggered susceptibility at
zero temperature away from the edges is more general than previously
 expected.
\begin{figure}[tbhp]
\centerline{\psfig{figure=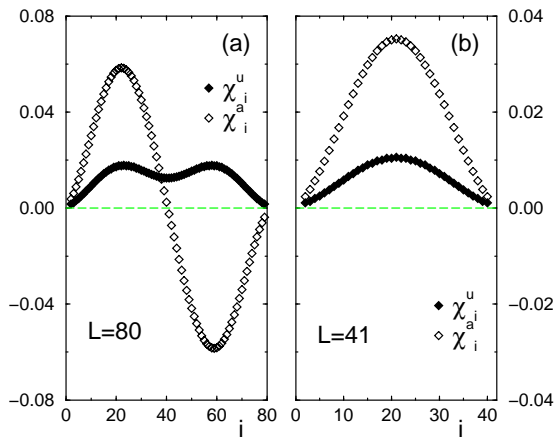,height=6cm,angle=270}}
\vspace{0.4cm}
\caption{Uniform and staggered components of the local susceptibility
$\chi_i$ obtained with the DMRG method applied to the frustrated $S=1/2$
Heisenberg model with $J_2/J_1 = 0.5$ on a chain with $L$ sites, using
$m$ states in the iterations and studying the subspace 
with a total spin
in the $z$-direction equal to 1.
(a) corresponds to $L=80$ and $m=32$, and (b) to $L=41$ and $m=32$.}
\end{figure}

Here it is important to remark that the behavior described in this
subsection could actually be observed in spin-Peierls systems such as
$CuGeO_3$. To understand this statement note that the similarities
between the results of subsections V.B and V.C are caused by the
dynamical adjustment in the latter of the pattern of spin dimers
that appear at $J_2/J_1= 0.5$. When a vacancy breaks a spin singlet,
the local ``damage'' is avoided by a dynamical process similar to the one
discussed in Fig.7. For an even number of sites between vacancies
this effect heals completely the damage (while with an
odd number of sites a soliton appears at mid-distance between the
vacancies). Thus, there are no spin-1/2 states localized near the
$Zn$-vacancies. A similar process is expected to occur in spin-Peierls
systems where  the phonons are dynamical variables which can also
adjust the pattern of strong and weak links. Preliminary 
Monte Carlo numerical results show that this effect indeed occurs in
practice~\cite{riera3}.

\subsection{$\chi_i$ in the dimerized Heisenberg chain}

The behavior for the case of a dimerized $J_1 - J_2 - \delta$ 
spin chain is quantitatively
 very different
from that observed in the $S=1/2$ Heisenberg model with and without
frustration, although qualitatively they are related. 
Fig.15 shows results for a representative value of the
dimerization $\delta = 0.048$~\cite{didier}. 
The first  and last links are ``weak'' in Fig.15.
Here $\chi_i^{u}$ is nonzero
only in the neighborhood of $i=1$ and $i=N$. By symmetry the spin near
each end is 1/2, since the total spin is 1 for an even number of sites
in the cluster.
It is important to note that
the staggered component is also localized  in approximately the
same range as the uniform one since the bulk susceptibility 
must vanish. These
results coexist with a spin-spin
enhancement of the static correlations which are also concentrated on a finite
region near the ends (Sec.II). 
\begin{figure}[tbhp]
\centerline{\psfig{figure=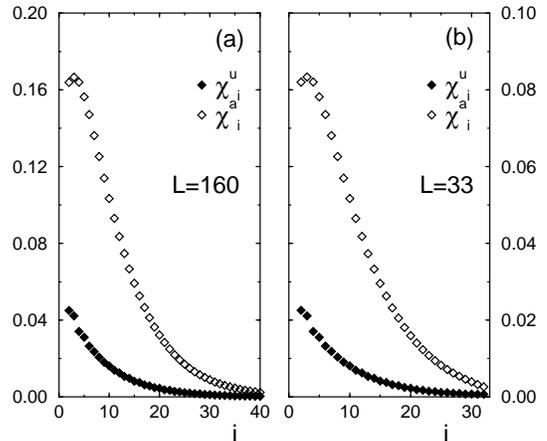,height=6cm,angle=270}}
\vspace{0.4cm}
\caption{Uniform and staggered components of the local susceptibility
$\chi_i$ obtained with the DMRG method applied to the dimerized $S=1/2$ chain
with no frustration $J_2/J_1 = 0$ and $\delta = 0.048$
on a chain with $L$ sites, using $m$ states in
the iterations and studying the subspace with a total spin
in the $z$-direction equal to 1.
(a) corresponds to $L=160$ and $m=20$, and (b) to $L=33$ and $m=32$.
In both cases the first link after the left chain end is ``weak''.}
\end{figure}
\noindent
The absence of spin-charge separation produces
virtually negligible size effects in the figure which is manifest in
the vanishing of both components at distances from the edges
 approximately larger than 40.
This has to be contrasted with Fig.13 for the undimerized case
where the staggered component
changed appreciably with the lattice size moving the location of its
maximum away from the edge. In the dimerized case and for the parameter
$\delta$ used here, the maximum is
at two lattice spacings from the chain end and remains there as the clusters
grow. 
Again, all this is a simple manifestation of the different 
localized vs delocalized character of the spin at the edges
for dimerized and undimerized chains, respectively.  Fig.15b
contains results for an odd number of sites $L=33$. With only a total
spin 1/2 to be distributed the chain selects one end as the depositary
of such spin and leaves the other spinless. This depends on whether
the first link on the chain, say starting from the left, is weak or
strong (the weak link carries the localized spin). 
Finally, note that if the first and last
links would have been ``strong'' then there is no reason for having
a localized spin 1/2 near the end. In this situation
 the spin is expected to
be delocalized with a maximum probability of being
located near the center of the chain (as illustrated below in Fig.16b).

Combining frustration and dimerization  
($J_2/J_1  = 0.35$ and $\delta =
0.012$) gives results similar to those of Fig.15. The localization length 
is only
slightly larger than for the case $J_2/J_1 = 0$, $\delta  = 0.048$.
Now the maximum in the staggered component is at site $i = 5$ 
which also reflects on a larger extension of the region where the
spin-spin correlations are perturbed with respect to their bulk behavior
by the influence of the chain ends. 

% since both frustration and
%dimerization are needed at the Hamiltonian level~\cite{riera1}. 
%The results for $\chi_i$ are shown in
%Fig.16. Once again it is observed that the uniform component is
%localized near the chain ends suggesting the presence of a state
%of spin 1/2 in their vicinity. Results for even and odd
%number of sites are very similar. 
%The localization length 

Note that for a short chain of
e.g. 20 sites, the staggered susceptibility would behave very similarly
to the results for the unfrustrated undimerized case Fig.12a. It is
clear that if the spin localization length is similar to the length of 
the chain (which mimics the distance between Zn impurities in real
systems) then the spin and charge are deconfined.
Also the qualitative trend of having a smaller disturbance in size
 near the ends as $\delta$ grows 
can be shown in the extreme case of a large dimerization such
as $J_2/J_1 = 0$, $\delta = 0.5$ (Fig.16a). Here the staggered
susceptibility reaches its maximum at $i=1$ i.e. in the first site
of the chain, and the uniform component is localized within 3 or 
4 lattice spacings of the chain end. 

Finally, in Fig.16b results are
presented for the case $J_2/J_1  = 0.35$ and $\delta =
0.012$ treated before,
but now with the first
and last links being ``strong''. The uniform susceptibility suggests
that the spin is not localized at the ends but it lives mainly
near the center as it occurs for a free particle in a square-well potential.
This agrees with the previous discussion where it was argued that
in a dimerized chain the $S=1/2$ near $Zn$ appears only on one side
of the vacancy, the one with the weak link immediately next to $Zn$.
Thus, qualitatively the results are as in the case of spin-charge separation
discussed before, and the staggered susceptibility has a behavior similar
to that of Fig.12a for the $S=1/2$ Heisenberg chain. Thus, assuming that
$Zn$ doping simply produces severed chains that keep their pattern of
weak and strong bonds 
(frozen phonons), there will be segments with even sites where
the spins are both
either localized or delocalized, depending on whether the first link
is weak or strong, and also segments with an odd number of sites where 
there is always
one localized spin. The latter was illustrated in Fig.12.b for the case
of a $L=33$ chain. 
\begin{figure}[tbhp]
\centerline{\psfig{figure=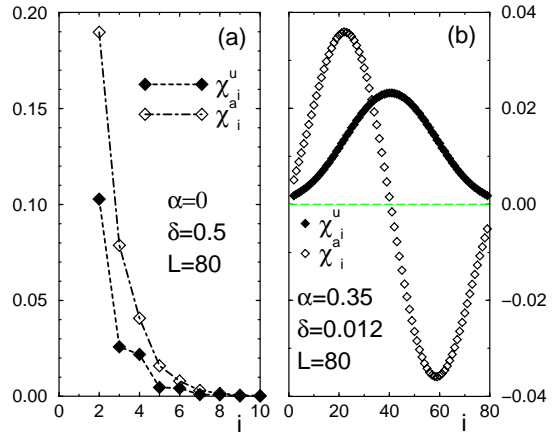,height=6cm,angle=270}}
\vspace{0.4cm}
\caption{Uniform and staggered components of the local susceptibility
$\chi_i$ obtained with the DMRG method applied to the dimerized $S=1/2$ chain
with frustration
on a chain with $L$ sites, using $m$ states in
the iterations and studying the subspace of total spin $z$-projection 1.
(a) corresponds to $\delta = 0.5$, $J_2/J_1 = 0.0$,
$L=160$ and $m=20$, with the first links being``weak'', 
and (b) has $\delta = 0.012$, $J_2/J_1 = 0.35$, $L=80$ and
$m=32$ with the first links being ``strong''.}
\end{figure}
Once again note that our discussion of subsection V.C
suggests that in the case of a compound where the coupling of spins and
lattice is dynamical, such as in the spin-Peierls systems, the pattern
of strong and weak links adjusts such that the first and last links
are always strong in a given chain segment with an even number of 
sites. Thus, the results of Fig.16b are relevant for this type of compounds.

\subsection{$\chi_i$ in a Heisenberg ladder cluster}

Similar calculations can be carried out  for ladder systems.
In Fig.17 results for the full local susceptibility $\chi_i$
are given for a ladder with $2 \times 32$ sites and vacancies
located at rung 11 and 22 in the same sublattice (two impurities
are introduced to produce an integer total spin). For the case
of no impurities, the susceptibility (which is entirely ``uniform'')
does not have structure showing that the ends of the ladder
do not localize spin
(remember that open boundary conditions are used in the DMRG method).
On the other hand, when the vacancies are introduced a clear enhancement
in $\chi_i$ is observed which is correlated with 
the position of the vacancy in, e.g., rung 11. 
The maximum change is always in the site located in
\begin{figure}[tbhp]
\centerline{\psfig{figure=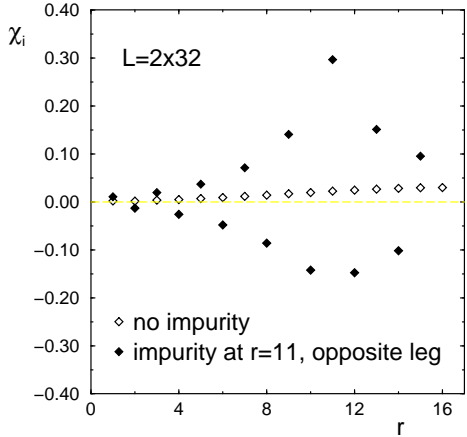,height=6cm,angle=270}}
\vspace{0.4cm}
\caption{Local susceptibility for the Heisenberg model on 
a $2 \times 32 $ ladder using DMRG,
keeping $m=16$ states, and studying the state of lowest energy in the
subspace of total spin in the $z$-direction equal to 1. Results with
and without impurities are shown, in the former running along the
opposite leg from those impurities which are located at rungs \mbox{11 and 22.}
}
\end{figure}
\noindent 
front of the vacancy along the same rung.
Fig.18a,b show the uniform and alternating components. 
\begin{figure}[tbhp]
\centerline{\psfig{figure=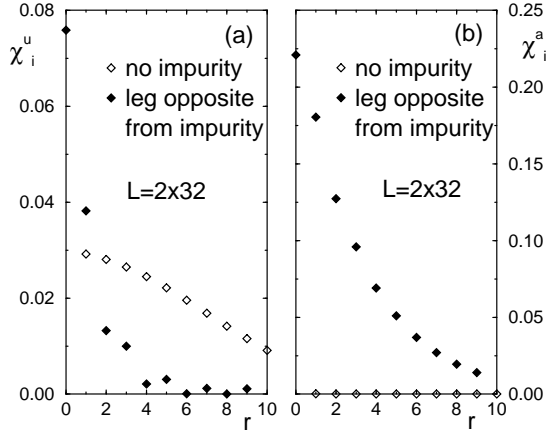,height=6cm,angle=270}}
\vspace{0.4cm}
\caption{Local susceptibility for the Heisenberg model on 
a $2 \times 32 $ ladder using DMRG,
keeping $m=16$ states, and studying the state of lowest energy in the
subspace of total spin in the $z$-direction equal to 1. Results with
and without vacancies are shown. (a) contains the uniform component and
(b) the staggered one. }
\end{figure}
The
localization length for the $S=1/2$ seems of about 2 to 3 lattice
spacings (as previously noticed by Sandvik et al.~\cite{sandvik}).
Then,
ladders and dimerized chains both behave very similarly
presenting localized $S=1/2$ states near vacancies and an enhanced
staggered susceptibility in their vicinity. On the other hand, the
Heisenberg model with and without frustration has no localized spins
near vacancies and a staggered susceptibility that grows with the
distance from the vacancy. In spite of these differences, 
below it is argued working at  finite
temperatures and/or high concentration of vacancies 
both families of models should present qualitatively similar
behavior that can be observed experimentally using NMR techniques.

\section{Monte Carlo results for dimerized chains}

To study the temperature dependence of some of the results
shown in previous sections, a standard world-line Monte Carlo simulation of
the dimerized unfrustrated \mbox{( $\delta \neq 0$, $J_2/J_1 = 0$)}
Heisenberg model on a chain was performed. 
The limitation of having $J_2/J_1 = 0$ is technical:
the addition of frustration
would have spoiled the simulation due to the appearance of a ``sign
problem''. 
%In spite of this
%restriction our results are  useful to compare theoretical predictions with
%the behavior of
%some real materials such as $Na V_2 O_5$ which do not seem to have
%frustration in its low energy effective spin Hamiltonian~\cite{didier}.
Results for the case of a $L=80$ chain, open boundary conditions,
$\delta = 0.05$, and with ``weak'' first and last links are shown in
Fig.19a,b (to collect good statistics about 3 to 4 independent runs
with $10^6$ sweeps each were carried out. The maximum number of
Trotter slices used was 120).
The behavior of the uniform susceptibility near the end is
particularly interesting. It shows the development of the spin 1/2
bound state as the temperature is reduced, in excellent agreement with
the DMRG and Lanczos results. The crossover from a
localized to delocalized spin 1/2 
occurs roughly
near $T^* \sim 0.1 J$, and this should be a temperature of relevance in
the experimental study of materials with structural dimerization.
%
%Since $J \sim 400K$ in $Na V_2 O_5$, then $T^* \sim 40 K$ is a
%relevant temperature for this material to observe the effects described
%in this paper. 
%
The staggered component follows
a similar pattern. 
\vspace{-0.4cm}

\begin{figure}[tbhp]
\centerline{\psfig{figure=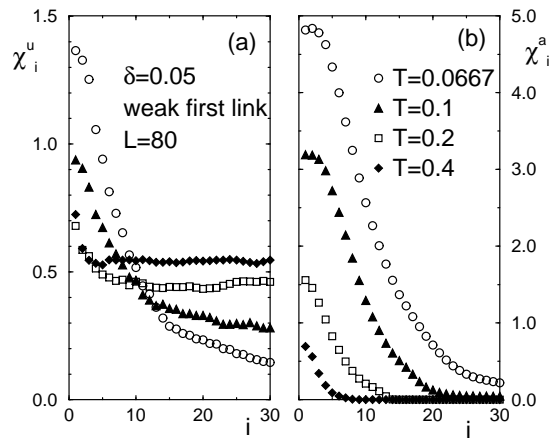,height=6cm,angle=270}}
\vspace{0.4cm}
\caption{Staggered susceptibility for the dimerized spin model
 obtained with Monte Carlo techniques on a chain with 80 sites
and OBC. The temperatures and $\delta$ are shown. (a) contains the
uniform component while (b) has the staggered one. The distribution of
dimerized links is such that the first link on the chain immediately
after the edge is ``weak'' (thus, generating a spin 1/2 localized state
in its vicinity at very low temperature).}
\end{figure}
\noindent
Fig.20.a,b illustrates what occurs in the case where the first and last links
are ``strong''. As discussed before no localized states are expected
in this situation and indeed the uniform susceptibility has no
structure near the end of the chain even at low temperatures.
Nevertheless it is curious that
the staggered component shows structure at intermediate
temperatures, although their intensity is much smaller than for the case
of ``weak'' links at the ends.

Finally, for completeness
in Fig.21a,b results for the pure $S=1/2$ unfrustrated undimerized
Heisenberg model are shown to compare Monte Carlo predictions with the DMRG results
of Fig.12. 
\begin{figure}[tbhp]
\centerline{\psfig{figure=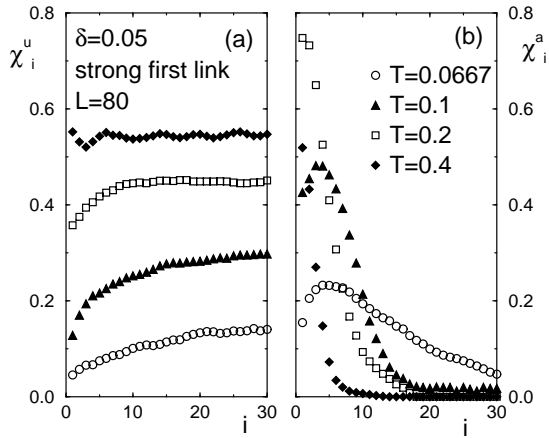,height=6cm,angle=270}}
\vspace{0.4cm}
\caption{Same as Fig.19 but now with a ``strong'' first link. In this
situation there are no localized spin 1/2 states near the edges.}
\end{figure}

\begin{figure}[tbhp]
\centerline{\psfig{figure=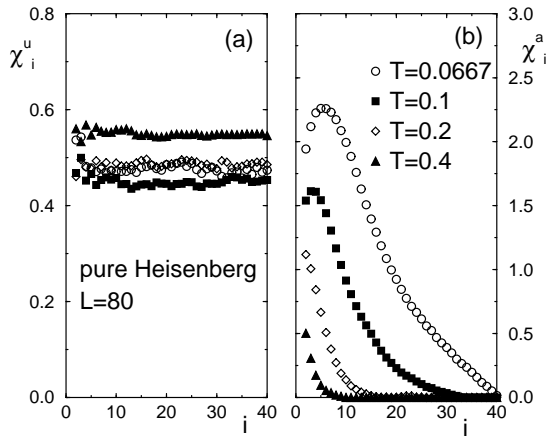,height=6cm,angle=270}}
\vspace{0.4cm}
\caption{Staggered susceptibility for the spin 1/2 Heisenberg  chain
obtained with Monte Carlo techniques on a 80 sites chain and using OBC.
Data for different temperatures are shown. (a) contains the uniform
component, while (b) has the staggered one.}
\end{figure}
\noindent
The uniform component has no structure as expected, while the
staggered one develops a strong peak which grows in intensity and moves
the location of its maximum away from the edge as the temperature is
reduced, in qualitative agreement with Fig.12.
Fig.22 contains results using the Monte Carlo approach, with and without
constraining the simulation to the subspace with a total 
$z$-component of the spin
equal to 1, and DMRG results in the lowest energy state of total
$z$-component spin 1, as used in most of Sec.V. 
\begin{figure}[tbhp]
\centerline{\psfig{figure=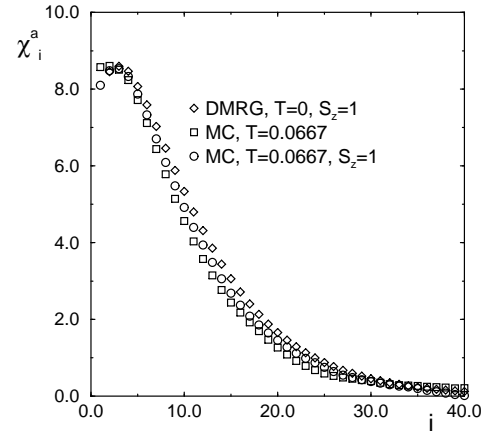,height=6cm,angle=270}}
\vspace{0.4cm}
\caption{Staggered susceptibility for the dimerized spin model obtained
using different techniques. The DMRG results were calculated on a
160 sites chain with $\delta = 0.048$, and a total $z$-component spin
in the ground state equal to 1. The Monte Carlo (MC) results were
calculated on chains with 80 sites and $\delta = 0.05$. In one case the
total $z$-component of the spin was fixed to 1, and in the other not.
The distribution of dimerized links is such that the first link on the
chain immediately after the edge is ``weak''. The amplitudes have been
adjusted such that the maxima have the same height. The agreement
between the different curves suggests that the methods of calculation
used in this paper are equivalent at low temperatures.}
\end{figure}
The excellent agreement
between all the results show that the three methods are quantitatively
equivalent at low temperatures. Similar results have been obtained in
the context of the 2-leg and 3-leg ladders by Sandvik et al.~\cite{sandvik}.

\section{Predictions for the NMR spectrum}

As explained in the introduction,
previous calculations in the context of the
$S=1/2$ Heisenberg chain showing that the local susceptibility
near a chain edge has a large alternating component led to interesting
predictions for the temperature dependence of some 
 NMR experiments~\cite{eggert}. 
More specifically, the alternating staggered
magnetization that such a local susceptibility would produce in the
presence of a magnetic field can be detected as a broadening of the
NMR spectra. These predictions have been recently summarized and
verified experimentally by Takigawa et al.~\cite{takigawa}. 
Let us denote as $K_i$ the shift at site $i$ of the resonance field 
induced by the hyperfine interaction between nuclear and electron spins.
The NMR spectrum simply represents the distribution function of $K_i$'s.
However, it can be shown~\cite{takigawa} that apart from a uniform shift,
$K_i$ is proportional to $\chi_i^{a}$. Actually 
$^{63}Cu$ NMR experiments carried out in $Sr_2 Cu O_3$, a compound 
with dominant one dimensional structures presumably well described by
the $S=1/2$ Heisenberg model, have provided results
compatible with the theoretical predictions~\cite{takigawa}.

The previous discussion shows that
the distribution of the local 
susceptibility or its staggered component
provides  interesting information for NMR experiments.
This information can be easily obtained from the results of
Sec.V. The qualitative discussion presented in this paper
shows that the effect observed before for the $S=1/2$ Heisenberg 
model~\cite{eggert}
is a particular
case of a more general situation. Our analysis predicts that 
if materials can be found that are a physical realization of frustrated
$S=1/2$ Heisenberg models ($J_2/J_1 \neq 0$) they would present NMR
spectra very similar to those of $Sr_2 Cu O_3$. In addition,
spin-Peierls systems may also have a NMR spectra  similar to that of
undimerized chains since phonons can adjust dynamically the pattern of
strong and weak links, avoiding the presence of localized spin-1/2
states near vacancies.
\begin{figure}[tbhp]
\centerline{\psfig{figure=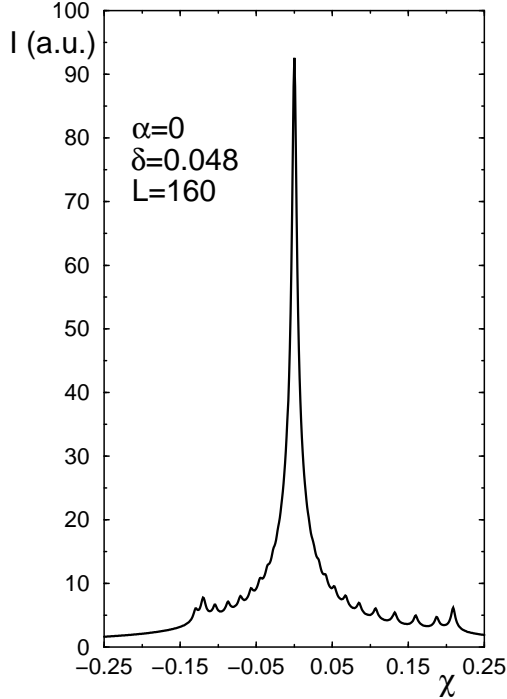,height=10cm,angle=0}}
\vspace{0cm}
\label{nmrplot}
\caption{NMR spectra for a dimerized chain with $\delta = 0.048$
using a ``weak'' first link on the chain and the DMRG technique (keeping
$m=20$ states).}
%The study is carried out in the lowest energy state with
%spin 1.
%The parameters used correspond to $Na V_2 O_5$.}
\end{figure} 

In addition, nontrivial 
results can be obtained also for structurally
 dimerized chains, such as  
$Cu(NO_3)_2\cdot2.5H_2O$, $CuWO_4$, $(VO)_2P_2O_7$, and 
$Sr_{14}Cu_{24}O_{41}$. In this case the
local susceptibility does $not$ grow with distance at zero temperature
due to the fact that there is no spin-charge separation in the system,
i.e. a spin 1/2 is localized near the chain ends. Nevertheless, a
broadening
in the overall NMR spectra is predicted as the temperature dependence
of the Monte Carlo results in
Fig.19 suggests. In this case, since the magnetic
susceptibility
in the bulk is zero, the distribution of $\chi_i$ will be centered
at zero. As $T$ is reduced the maximum value of $|\chi_i|$, related with
the total width of the NMR spectra, moves away from the
chain end, as the MC results of Fig.19b imply. The central peak must increase
its width as the temperature is reduced.

%To make this conclusion more explicit,
%results are shown in Fig.24 containing the temperature
%evolution
%of the distribution of $\chi_i$'s
%using the dimerized Heisenberg model with parameters compatible with
%the compound $Na V_2 O_5$. The broadening discussed before is clearly
%observed. 
Results for the distribution of $\chi$'s
 obtained with DMRG techniques at zero temperature
but, as discussed before, 
in the total spin projection $S_z=1$ subspace are presented in
Fig.23.
% for parameters compatible with
%$Na V_2 O_5$. 
They qualitatively agree with the results of the Monte Carlo
simulations at the lowest temperature available. The actual temperature
dependence of these MC results is given in Fig.24.
\begin{figure}[tbhp]
\centerline{\psfig{figure=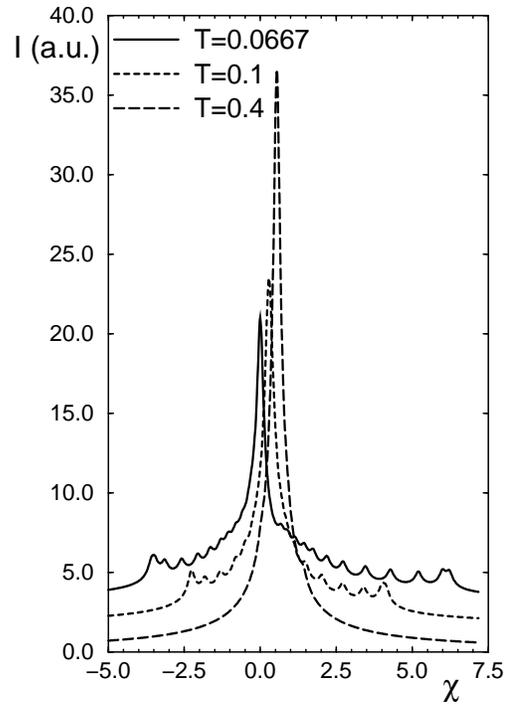,height=10cm,angle=0}}
\vspace{0cm}
\caption{Same as Fig.23 but now calculated with Monte Carlo simulations at
different temperatures (in units of $J$) using the results of Fig.19.
The broadening of the shoulder with lower temperatures reflects the
increase of the local susceptibility at the boundaries. The shift of the
main peak with increasing temperature is due to the increased ``bulk''
susceptibility.}
\end{figure} 
\noindent

Fig.25a contains
DMRG results for $J_2/J_1= 0.35$ and $\delta = 0.012$.
In this case Monte
Carlo simulations cannot be performed due to technical reasons (sign problems).
Nevertheless, the DMRG result is qualitatively similar
to those in Fig.23.
The NMR spectra contained in Figs.23-25a  
are  predictions of our calculations
that could be checked experimentally to address the universality
of the staggered susceptibility enhancement near chain ends
described in this paper.
Qualitatively the predictions are similar as reported by Sandvik et
al.~\cite{sandvik} in the context of the 2-leg and 3-leg spin ladders.

Note that the results of Fig.25, as well as the previous
NMR calculations~\cite{eggert},
have been performed on a single chain of length $L$,
with $L$ large. These results can only be applied to real systems
with a very low concentration of impurities. Actually the experiments
in $Sr_2 Cu O_3$~\cite{takigawa} were carried out without adding explicit doping but simply
expecting the spontaneous presence 
of a finite (and very small) concentration of vacancies in the system. 
In the same spirit, experiments in structurally dimerized chains
or spin-Peierls systems can be conducted
also without $Zn$ doping, and the results should be close to those
discussed in Figs.23 and 25b. 
\begin{figure}[tbhp]
\centerline{\psfig{figure=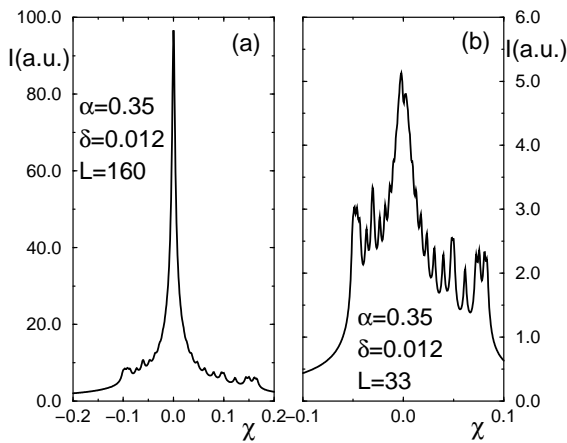,height=6cm,angle=270}}
\vspace{0.4cm}
\caption{(a) Same as Fig.23 but using $J_2/J_1=0.35$, and $\delta = 0.012$;
%These parameters mimic the behavior of $CuGeO_3$; 
(b) NMR spectra for
the case of a ``short'' chain of 33 sites. $m=32$ states were kept in the
DMRG technique.}
\end{figure}
\noindent
However, 
for completeness, here results for short chains are also shown
to mimic the behavior of structurally
dimerized compounds explicitly doped by $Zn$. 
For instance, if the length of the chain is $L=33$, then the results should be
contrasted with $Zn$-doped $x \sim 0.03$.
A representative result is shown in Fig.25b. Note
that now the central peak is much broader than in Fig.25a,
and the background has comparably much more intensity. Results for
even shorter chains reveal the discrete nature of the chain, presenting
just a small number of peaks in the hystogram. 
However, in a real system the
Zn impurities will not lie uniformly distributed at distance
$l\sim 1/x$, but they will follow a random pattern producing a
distribution of chain lengths that includes short and long ones.
(Fig.26 shows explicitly the distribution of distances between two
impurities). Thus,
it is reasonable to expect experimentally
a mixing of results such as those of Figs.25a and 25b for $Zn$-doped compounds.
In spite of these complications, for small values of $x$ 
clearly a large central peak will appear as a dominant
structure on top of a background that increases its size (without
diverging for dimerized systems) as the temperature is reduced.

Very recently NMR spectra for $Zn$-doped $SrCu_2 O_3$, which is 
a spin ladder system, has been presented~\cite{fujiwara}. The results show
a large broadening of the main signal in the NMR spectra as the temperature
is reduced from 280K to 20K for the case of a $Zn$ concentration of just
$0.25\%$ and $0.50\%$. 
\begin{figure}[tbhp]
\centerline{\psfig{figure=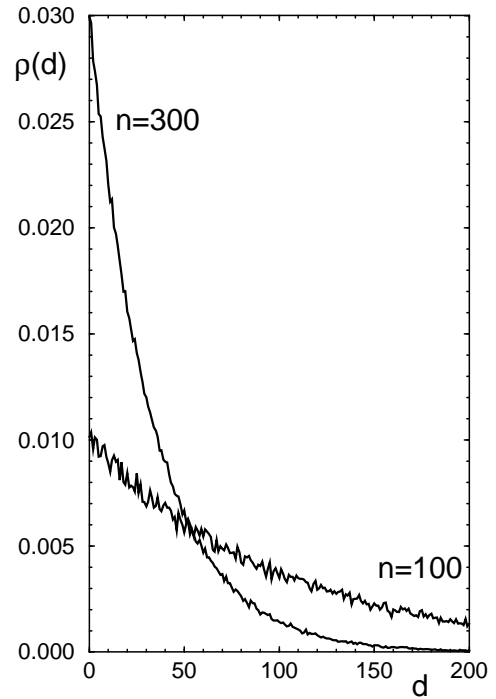,height=10cm,angle=0}}
\vspace{0cm}
\caption{Distribution of distances d between two impurities working on a
chain of length $10,000$ sites, with $n$ impurities randomnly
distributed.
Each curve is an average over $1,000$ distributions of impurities.}
\end{figure}
\noindent
The authors conclude that a picture where each
doped vacancy induces a spin 1/2 localized in the same rung as the vacancy
cannot explain this result since the vast majority of the spins would
be unaffected by doping. This conclusion is in excellent agreement with the
observation throughout this paper that for realistic values of parameters
the region in the vicinity of the vacancies where the staggered local
susceptibility is enhanced can be large,  involving several lattice
spacings. 
%For the case of $Na V_2 O_5$, where
%both DMRG and Monte Carlo results are available (Figs.15 and 19), it is clear that
%about 30 sites near the edge of the open BC chain contribute to the 
%staggered susceptibility. 
For ladders Fig.18 suggest that in this case
$\sim 10$ rungs (i.e. 20 sites) away from the vacancy
still have enhanced antiferromagnetism. If a 2-leg ladder
 with $2 \times 100$
sites and open BC is considered
(to  mimic the undoped ladder segments for
a nominal $Zn$-concentration of $0.5 \%$), then $20\%$ of the system
is affected by the edges. In addition, the extra broadening effect caused
by the probabilistic existence of short segments (see Fig.26) 
should in principle be
taken into account to contrast theory with experiments. This detailed
calculation will be carried out in future publications, but here it is safe 
to conclude that the recent NMR spectra in ladder systems is compatible
with the results presented in this paper.

\section{conclusions}

The effect of vacancies on a variety of spin models and
geometries has been studied with the help of computational techniques.
Near these vacancies the staggered
 spin-spin correlations are clearly enhanced. The range of the effect
can vary from just a few lattice spacings (for strongly dimerized
systems or models with large spin gaps where a spin 1/2 state
is bounded to the chain edge) to the whole lattice (for
the $S=1/2$ Heisenberg model with and without frustration   where there
is no spin 1/2 bound state near the vacancies). 
The origin of this phenomenon was here described in terms of
short-range RVB states. Introducing vacancies prunes or reduces
the number of possible spin singlet configurations and this enlarges
the spin correlations as described in this paper. The phenomenon
also takes place in 2D models, although the effect here is not
as large as in 1D. 
The AF enhancement
can be observed in the behavior of NMR experiments
as a broadening of the spectra as the temperature is reduced
in structurally dimerized chains, ladders such as
 $Sr Cu_2 O_3$, spin-Peierls compounds, and other materials
after $Zn$ impurities are introduced.
In addition, it is 
likely that the enlarged AF correlations can stabilize
3D N\'eel order upon $Zn$ doping
 once a weak correlation between chains and ladders
is incorporated, in agreement with experimental results~\cite{azuma,hase1}.

\section{acknowledgments}

M.L. and E.D. are supported by grant NSF-DMR-9520776. 
G.B.M. acknowledges the financial support of the Conselho
Nacional de Desenvolvimento Cient\'{\i}fico e Tecnol\'{o}gico (CNPq-Brazil).
C. J. G. acknowledges the financial support of the Consejo Nacional de
Investigaciones Cient\'{\i}ficas y T\'{e}cnicas (CONICET-Argentina).
Additional 
support by the National High
Magnetic Field Lab is acknowledged.


\begin{references}
\bibitem{comment2} Actually the 1-2 singlet is more probable
than 1-3 and 1-4  due to the characteristics of the 2-leg ladder
ground state~\cite{ladder}. 
This detail is not qualitatively  relevant here.
\end{references}
\end{document}